\documentclass[a4paper,12pt]{article}
\usepackage[pctex32]{graphics}

\newcommand{\be}{\begin{equation}}
\newcommand{\ee}{\end{equation}}
\newcommand{\ben}{\begin{eqnarray}\displaystyle}
\newcommand{\een}{\end{eqnarray}}

\begin{titlepage}

\begin{document}
{\baselineskip20pt

\vskip .6cm

\begin{center}
{\Large \bf Entropy bound of local quantum field theory with
generalized uncertainty principle}

\end{center} }

\vskip .6cm
 \centerline{\large Yong-Wan Kim$^{a}$, Hyung Won Lee$^{b}$, and Yun Soo Myung$^{c}$}

\vskip .6cm

\begin{center}
{Institute of Basic Science and School of Computer Aided
Science,\\Inje University, Gimhae 621-749, Korea }
  \end{center}
\renewcommand{\thefootnote}{\arabic{footnote}}
\setcounter{footnote}{0} \setcounter{page}{2}

\centerline{\bf Abstract} \bigskip
\begin{abstract}
  We study the entropy bound for local quantum field theory (LQFT) with generalized uncertainty principle.
The generalized uncertainty principle provides naturally a UV
cutoff to the LQFT as gravity effects.
  Imposing the non-gravitational collapse condition as the UV-IR relation,
  we find that the maximal entropy of a bosonic field is limited  by the  entropy bound $A^{3/4}$ rather than
  $A$ with $A$ the boundary area.

\end{abstract}


\vskip 0.8cm

\baselineskip=18pt
\noindent $^q$ywkim65@gmail.com \\
\noindent $^b$hwlee@inje.ac.kr\\
\noindent $^c$ysmyung@inje.ac.kr

\noindent
\end{titlepage}

\setcounter{page}{2}

\section{Introduction}
   The counting of degrees of freedom of local quantum field theory (LQFT) including gravity effects
is an important issue.
   For example, statistical mechanics  implies
     that a thermal photon gas could be  described by LQFT without gravity. Its entropy
    takes the form of $S \sim T^3L^3$  when it is confined to a box of size $L$ as an IR cutoff.
    If the temperature $T$ could be an arbitrarily chosen parameter,
      one finds that  the system has an entropy proportional to the volume $L^3$.
       However,  this temperature has to be limited by the energy bound $E \sim T^4L^3 \le E_{BH}\sim L$,
       or else the system will undergo collapse to form a black hole when considering the gravity effects.
       Applying this energy bound to the entropy,
        one finds the entropy bound of  $S \le S_{\rm max}\sim A^{3/4}$,
        where $A\sim L^2$ is the boundary area of the system.
        The derivation above is firstly given by 't Hooft in~\cite{tH}.
        The entropy bound $A^{3/4}$ for LQFT  was  also discussed  by other authors~\cite{CKN,BH,Myung,HR}.

   However, there are still controversies over this topic. Starting from a bosonic field model
    and imposing the gravitational stability condition,
a holographic entropy bound of $ S \le S_{\rm max} \sim A$ appears as
the covariant entropy bound. This may not describe the local
quantum field theory including  gravity effects because the  LQFT
requires a more stricter entropy bound rather
       than the holographic entropy bound.

In order to obtain a compact relation from the energy and entropy
bounds,  the  UV cutoff $\Lambda$ is necessarily introduced to
regularize the LQFT~\cite{CKN}. Explicitly, the LQFT with $E_{\Lambda}\simeq
\Lambda^4L^3$ and $S_{\Lambda}\simeq \Lambda^3L^3$ is  able to
describe a thermodynamic system at temperature $T$, provided that
$T\le \Lambda$. If $T \gg 1/L$, the energy and entropy will be
those for a thermal photon gas: $E_{R} \simeq T^4L^3$ and $S_{R}
\simeq T^3L^3$. In this case, the modes with momentum more than
the UV cutoff have been excluded from consideration. However, this
cutoff could be justified in an average sense,  and hence the
states with momentum $p>\Lambda$ should be accounted properly.

         On the other hand, one believes that the generalized uncertainty principle (GUP) arises from
         the Heisenberg uncertainty principle when gravity effect
         is taken into account~\cite{GUPs,GUPb,GUPg,GUPo}.
         Its  commutation relation of
         \begin{equation} \label{crgup}
         [\vec{x}, \vec{p}]=i\hbar(1+\beta^2\vec{p}^2)
         \end{equation}
         leads to the  generalized uncertainty relation~\cite{CMOT}
         \begin{equation} \label{GUP}
         \Delta x \geq \frac{\hbar}{2}\left(\frac{1}{\Delta p}+\beta\Delta
         p\right)
         \end{equation}
which  implies the presence of  a minimal length scale
          \begin{equation}
          \Delta x_{\rm min}=\hbar\sqrt{\beta}.
         \end{equation}
          Thus, the leading order   correction to the standard
          formula is expected to be proportional to the Planck length $l_P$,
if one chooses
          $\beta=G/c^3\hbar$ with the Newton's constant $G=c^3l_{P}^2/\hbar$.
Hence the GUP plays the same role as the UV cutoff $\Lambda$ does
show because the UV cutoff also determines a minimal detectable
length.

   In this Letter,  we  show that the entropy bound for LQFT is
   $A^{3/4}$ rather than $A$ by using the GUP and non-gravitational collapse condition,
    instead of the UV cutoff introduced by hand.

\section{State counting with UV cutoff }

We consider  a massless scalar field confined to  three-dimensional spacelike
cube of size $L$ in Minkowski space, as has been done  in~\cite{Yurt,Aste04,Aste06,CX}.
 The modes of the field are then the solution
to the wave equation $\nabla^2\Phi=0$ with periodic boundary conditions. Any mode of quantized wave vector
$\vec{k}$ could be labelled by three
positive integers $\vec{m}=(m_x,m_y,m_z)$ as $\vec{k}=\frac{\pi}{L}\vec{m}$.
The corresponding energy of the  mode is
\begin{equation}
E_{\vec{m}}=\hbar \omega_{\vec{m}}=\hbar c|\vec{k}|=
\frac{\hbar\pi c}{L}\sqrt{\vec{m}\cdot\vec{m}}\end{equation} where
we used the relation of $\omega=ck=p/\hbar$. Hereafter we use the
Planck units of $G=\hbar=c=k_B=1$, which implies a simple relation
of $p=\omega=k$. The total number of these quantized modes is
calculated by the replacement
\begin{equation}
 N=\sum_{\vec{k}}1 \rightarrow \frac{L^3}{(2\pi)^3}\int d^3\vec{k}
  =\frac{L^3}{2\pi^2}\int_{0}^{\Lambda}\omega^2 d\omega
  =\frac{\Lambda^3L^3 }{6\pi^2},
\end{equation}
where $\Lambda$ is introduced to be the UV energy cutoff of the
LQFT. See Fig.1 for the uniform distribution up to
$\omega=\Lambda$. Thanks to the UV cutoff, $N$ is finite and thus
there exists one-to-one correspondence between the wave vector
$\vec{k}$ and a character $i$ with $i\in[1,N]$. Upon quantization
of a massless scalar field which obey the Bose-Einstein
statistics, we can construct the Fock states by assigning
occupying number $n_i$ to these $N$ different modes
\begin{equation}
 \mid\Psi\!>=\mid n(\vec{k}_1), n(\vec{k}_2),\cdots,n(\vec{k}_N)>
 ~~\to~~ \mid n_1,n_2,\cdots,n_N>,
\end{equation}
where the normalized state contains $n(\vec{k}_1)$ particles with
momentum $\vec{k}_1$, $n(\vec{k}_2)$ particles with momentum
$\vec{k}_2$, and so on. Thus, the basis of the Hilbert space
${\cal H}$ of the system is spanned by each different set of
$\{n_i\}$, and the number of occupancies $\{n_i\}$ gives the
corresponding dimension of the Hilbert space (${\rm dim} {\cal
H}$). Usually, the dimension of the Hilbert space is infinite for
bosons  unless the number of  particles in each mode $i$ is
constrained  by a finite bound. However, the non-gravitational
collapse requirement makes  the  permissible dimension of Hilbert
space finite as
\begin{equation} \label{gsc}
E= \sum_{i=1}^N n_i \omega_i  \leq
 E_{BH}=L.
\end{equation}
The number of solutions or occupancies $\{n_i\}$ satisfying the
above bound gives the dimension  $W \equiv {\rm dim} {\cal H}$ of
physically  permitted Hilbert space. In other words, in order to
determine the dimension of Hilbert space, one has to know the
number of admissible solutions $\{n_i\}$ satisfying
Eq.(\ref{gsc}). This corresponds to the knapsack or counting
lattice points problem. When confining $\{n_i\}$ to a Cartesian
coordinate system $\{x_i\}$, the question refers to the counting
lattice points contained within the convex polytopes determined by
\begin{equation}
\sum^N_{i=1}x_i\omega_i \le E_{BH}, ~~x_i\ge 0
\end{equation}
with right-angle side lengths
\begin{equation}
L_i = \frac{ E_{BH}}{\omega_i}, ~~{\rm with}~i\in[1,N].
\end{equation}
Actually, it is difficult to find an exact solution to this question.
For $L_i \gg1$, one may use the volume of the corresponding polytopes to approximately evaluate
the number of lattice points within them.

We note that  an $N$ particle state with one particle occupying
one mode ($n_i=1$) corresponds to the lowest energy state with $N$
modes simultaneously excited.
In this case, it should satisfy the
gravitational stability condition of Eq.(\ref{gsc}).
Hence, the energy bound is given by
\begin{equation}
E \rightarrow
\frac{L^3}{2\pi^2}\int_{0}^{\Lambda} \omega^3 d\omega =\frac{\Lambda^4 L^3}{8\pi^2} \leq
 E_{BH}.
\end{equation}
The last inequality implies the UV-IR relation
\begin{equation} \label{UVR}
\Lambda^2\leq \frac{1}{L}.
\end{equation}
On the other hand, the entropy associated with the system is given
by
\begin{equation}
S=-\sum^{W}_{j=1} \rho_j \ln \rho_j,
\end{equation}
where $\rho_j$ is the possible distribution of the Hilbert state
basis. It is clear that the maximum value of the entropy  is
realized by taking a uniform distribution of $\rho_j=1/W$. Then,
the maximum entropy is given by
\begin{equation}
S_{\rm max}=-\sum_{j=1}^W \frac{1}{W}\ln\frac{1}{W},
\end{equation}
where the bound of $W$ is determined  by
\begin{equation}
W ={\rm dim}{\cal H}< \sum^N_{m=0}\frac{z^m}{(m!)^2} \le
\sum^\infty_{m=0}\frac{z^m}{(m!)^2}=I_0(2\sqrt{z}) \sim
\frac{e^{2\sqrt{z}}}{\sqrt{4\pi\sqrt{z}}}.
\end{equation}
Here $I_0$ is the zeroth-order Bessel function of the second kind.
 Since $z$ is given
by
\begin{equation}
 z= \sum^N_{i=1}L_i \to  \frac{L^3}{2\pi^2}\int_{0}^{\Lambda}\Bigg[\frac{E_{BH}}{\omega}\Bigg]\omega^2d\omega
 =\frac{\Lambda^2L^4}{4\pi^2},
\end{equation}
we find the bound
\begin{equation}
z  \leq  L^3,
\end{equation}
where we used the UV-IR relation in Eq.(\ref{UVR}). Therefore, we
have the bound for the maximum entropy
\begin{equation}
S_{\rm max}=\ln W \leq A^{3/4}.
\end{equation}
This is a brief derivation of the entropy bound by using the LQFT.
\begin{figure}[t!]
   \centering
   \includegraphics{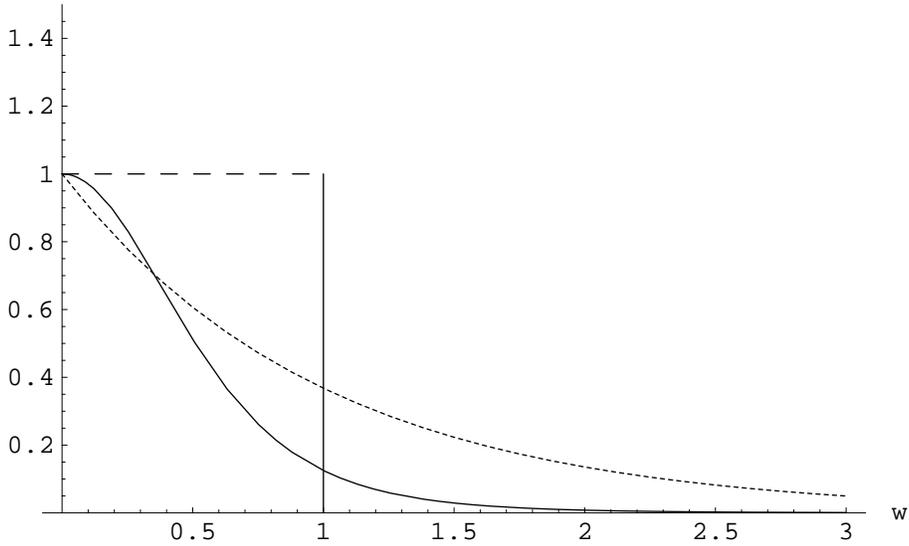}
\caption{Density functions for regularization of LQFT as function
of $\omega$. The dashed line denotes ``uniform density function"
as the UV cutoff, while the solid curve represents the GUP density
function in Eq.(\ref{dgup}) with $\beta=1$. If one considers the
solid curve within the box, it provides the density function for
UV cutoff and GUP for Section 4. The dotted curve denotes density
function for an all-order result in Appendix. Here we choose
$\Lambda=1/\sqrt{\beta}=1$.} \label{fig1}
\end{figure}

\section{State counting with GUP}

The GUP relation of Eq.(\ref{GUP}) has an effect on the density of
states in momentum space \cite{CMOT} as
\begin{equation} \label{dgup}
\frac{d^3\vec{p}}{(1+\beta \vec{p}^2)^3}
\end{equation}
with an important factor of $1/(1+\beta \vec{p}^2)^3$, which
effectively cuts off the integral beyond $p=1/\sqrt{\beta}$.
Intuitively, this can be understood from the observation that the
right-hand side of  Eq.(\ref{crgup}) includes a
$\vec{p}$-dependent term  and thus affect the cell size in phase
space as ``being $\vec{p}$-dependent". Rigorously, making use of
the Liouville theorem, one could show that the invariant weighted
phase space volume under time evolution is given by \cite{CMOT}
\begin{equation}
\frac{d^3\vec{x}d^3\vec{p}}{(1+\beta \vec{p}^2)^3},
\end{equation}
where the classical commutation relations corresponding to the
quantum commutation relation of Eq.(\ref{crgup}), as $\{x_i,
p_j\}=(1+\beta p^2)\delta_{ij}$, $\{p_i,p_j\}=0$, and
$\{x_i,x_j\}=2\beta(p_ix_j-p_jx_i)$, are used. This factor plays a
role of the UV cutoff of the consequent momentum integration, as
shown in Fig. 1.

For a massless scalar field confined to three-dimensional
spacelike cube of size $L$, the total number of modes can be
calculated as
\begin{equation} \label{ngup}
 N=\frac{L^3}{2\pi^2}\int_{0}^{\infty}\frac{\omega^2d\omega}{(1+\beta \omega^2)^3}
 =\frac{1}{32\pi}  \frac{L^3}{\beta^{3/2}} \sim \frac{L^3}{\beta^{3/2}}.
\end{equation}
 As expected, due to
strong suppression of density of state at high momenta, the total
number is rendered finite with $1/\sqrt{\beta}$ acting effectively
as the UV cutoff. This result is in strong contrast to the
previous calculations where the UV cutoff $\Lambda$ is an
arbitrary scale which must be introduced by hand, and where one
must assume that the physics beyond this cutoff does not
contribute (See Fig. 1).

The energy bound is given by
\begin{equation}
E \to
\frac{L^3}{2\pi^2}\int_{0}^{\infty}\frac{\omega^3d\omega}{(1+\beta
\omega^2)^3}
 =
 \frac{1}{8\pi^2}\frac{L^3}{\beta^2}\leq E_{BH}.
\end{equation}
The last inequality implies the important UV-IR relation
\begin{equation} \label{guv}
L\leq \beta.
\end{equation}
In order to calculate the maximum entropy, we need to know  $z$
which  is  calculated to be
\begin{equation}
 z \to  \frac{L^3}{2\pi^2}\int_{0}^{\infty}\Bigg[\frac{E_{BH}}{\omega}\Bigg]\frac{\omega^2d\omega}{(1+\beta \omega^2)^3}
 =
 \frac{1}{8\pi^2}\frac{L^4}{\beta}.
\end{equation}
Using Eq.(\ref{guv}), one has the bound for $z$ as
\begin{equation}
z \leq L^3 .
\end{equation}
Finally, we arrive at the same bound for the maximum entropy
\begin{equation}
S_{\rm max}=\ln W \leq A^{3/4}.
\end{equation}

\section{State counting with UV cutoff and GUP}

For a massless scalar field confined to  3-dimensional spacelike
cube of size $L$, the  mode counting method is  changed to include
the GUP effect (See the solid curve in the box in Fig.1)
\begin{equation}
 N \to \frac{L^3}{2\pi^2}\int_{0}^{\Lambda}\frac{\omega^2d\omega}{(1+\beta \omega^2)^3}
 \approx
 \frac{L^3}{2\pi^2}\left(\frac{\Lambda^3}{3}-\frac{3\Lambda^5\beta}{5}\right).
\end{equation} That is, the total number of states is  decreased when we
include the GUP effect.

The energy bound is modified up to $\beta$
\begin{equation}
 E \to \frac{L^3}{2\pi^2}\int_{0}^{\Lambda}\frac{\omega^2d\omega}{(1+\beta p^2)^3}
 \approx
 \frac{L^3}{2\pi^2}\left(\frac{\Lambda^4}{4}-\frac{\beta\Lambda^6}{2}\right)\nonumber\\
 \leq E_{BH}.
\end{equation}
The last inequality implies
\begin{equation} \label{buvg}
\Lambda^2\leq \frac{1}{L}\left(1+\frac{\beta}{L}\right).
\end{equation}
The maximum entropy is given by
\begin{equation}
S_{\rm max}=-\sum_{j=1}^W \frac{1}{W}\ln\frac{1}{W},
\end{equation}
with  $W\sim e^{2\sqrt{z}}$. Since $z$ is given  up to $\beta$ by
\begin{equation}
 z \to  \frac{L^3}{2\pi^2}\int_{0}^{\Lambda}\Bigg[\frac{E_{BH}}{\omega}\Bigg]\frac{\omega^2d\omega}{(1+\beta \omega^2)^3}
 \approx
 \frac{L^4}{2\pi^2}\left(\frac{\Lambda^2}{2}-\frac{3\beta\Lambda^4}{4}\right),
\end{equation}
one finds the bound when using Eq.(\ref{buvg})
\begin{equation}
 z \leq  L^3 -\frac{\beta L^2}{2}.
\end{equation}
Therefore, we have a modified bound for the maximum entropy
\begin{equation}
S_{\rm max}=\ln W \leq A^{3/4}-\frac{\beta}{4}A^{1/4}
\end{equation}
which shows clearly that the upper bound is decreased due to the
GUP.

\section{Discussions}
In this work, we show how gravity effects offer a way to calculate
the maximal entropy bound of the LQFT. Gravity effects provide UV
and IR cutoffs to the LQFT.  If the GUP is really considered as a
reflection of gravity effects, it gives a UV cutoff which makes
the total number of modes $N$ finite.  On the other hand, the
energy bound implies that the number of particles $n_i$ is limited
and the total energy of the system is less than that of the
same-sized black hole. The former makes the dimension of Hilbert
space finite, while the latter leads to the bound as the UV-IR
relation.  Then, we obtain the maximal entropy bound of $A^{3/4}$.
This is consistent with the picture that the gravity effects make
the dimension of Hilbert space finite.

We wish to emphasize why our work is meaningful by comparing it
with two known approaches. Without the UV cutoff, we could derive
the entropy bound of $S \le S_{\rm max} \sim A^{3/4}$, as
suggested by 't Hooft in~\cite{tH},  if the UV cutoff is much
larger than the temperature and the energy bound is imposed.
However, this corresponds to a heuristic derivation  because the
LQFT was not explicitly used for calculation. As was briefly
reviewed in Sec. 2, we introduce UV and IR cutoffs to calculate
the entropy bound  of a massless field when using the LQFT. This
means that we need  both UV-control and IR-control to get an
important UV-IR relation of Eq.(\ref{UVR}). However, as is shown
Fig. 1, the UV cutoff $\Lambda$ may correspond to an {\it ad hoc}
density function because it was introduced by hand.  In this work,
we introduce the new cutoff $\beta$ based on the GUP which
effectively cuts off the short distance region. This case provides
a more natural derivation of $A^{3/4}$ than using the UV cutoff
$\Lambda$ because the GUP is considered as a meaningful  extension
of the first principle ``Heisenberg uncertainty principle" when
taking into gravity effects account.

We mention that as was shown in Eq.(\ref{ngup}), the total number
of modes $N$ is  clearly determined by imposing the GUP. In order
to confirm this, we introduce  an all-order result to the
Heisenberg uncertainty relation. As is shown in Appendix, we
choose the commutation relation as a  way of implementing
all-order GUP corrections. Then, the total number $N$ of modes in
Eq.(\ref{alln}) takes a similar form as in Eq.(\ref{ngup}). This
supports that the GUP corrections to the Heisenberg uncertainty
relation is equivalent to a UV cutoff to the LQFT.

 Finally, we note that the
non-gravitational collapse condition plays the important role: it
makes the dimension of Hilbert space finite and thus provides the
bound of maximal entropy.

\section*{Appendix: An all-order result in GUP}
The GUP commutation relation in Eq.(\ref{crgup}) can be extended
into~\cite{Nou}

         \begin{equation}
         [\vec{x}, \vec{p}]=i\hbar e^{\beta^2\vec{p}^2},
         \end{equation}
which  includes all order corrections to the Heisenberg
uncertainty principle. In this case, the weighting factor is given
by~\cite{KP}
\begin{equation} \label{allgup}
d^3\vec{p}~ e^{\beta^2\vec{p}^2}.
\end{equation}
The total number of modes is calculated to be
\begin{equation} \label{alln}
N \to \frac{L^3}{2\pi^2}\int^{\infty}_{0} \omega^2
e^{-\beta^2\omega^2} d\omega =
\frac{L^3}{8\pi^{3/2}\beta^{3}}\sim\frac{L^3}{\beta^{3}}
\end{equation}
without any ambiguity.
 The energy bound is obtained as
\begin{equation} \label{alleb}
E\to \frac{L^3}{2\pi^2}\int^{\infty}_{0} \omega^3
e^{-\beta^2\omega^2} d\omega = \frac{L^3}{4\pi^2\beta^4}\le
E_{BH}=L,
\end{equation}
where the last equality implies the UV-IR relation as the bound
\begin{equation}
L\le\beta^2
\end{equation}
In order to compute the maximal entropy, one has to know the
variable $z$
\begin{equation}
z\rightarrow\frac{L^3}{2\pi^2}\int^{\infty}_{0}
\left[\frac{E_{BH}}{\omega}\right]\omega^2 e^{-\beta^2\omega^2}
d\omega = \frac{L^4}{4\pi^2\beta^2}.
\end{equation}
Using the energy bound of Eq.(\ref{alleb}), one finds  the bound
for $z$ as
\begin{equation}
z\le L^3.
\end{equation}
Thus, the maximum entropy bound is confirmed to be
\begin{equation}
S_{\rm max}=\ln W \leq A^{3/4}.
\end{equation}

\end{document}